\documentclass[nohyperref]{article}

\usepackage{microtype}
\usepackage{graphicx}
\usepackage{subfigure}
\usepackage{booktabs} % for professional tables

\graphicspath{ {./images/} }

\usepackage{hyperref}

\usepackage[accepted]{icml2022}

\usepackage{amsmath}
\usepackage{amssymb}
\usepackage{mathtools}
\usepackage{amsthm}

\usepackage[capitalize,noabbrev]{cleveref}

\theoremstyle{plain}

\theoremstyle{definition}

\theoremstyle{remark}

\newtheorem{hyp}{Hypothesis}

\usepackage[textsize=tiny]{todonotes}

\usepackage{enumitem}

\newlist{questions}{enumerate}{2}
\setlist[questions,1]{label=RQ.,ref=RQ}

\icmltitlerunning{Training Novices: The Role of Human-AI Collaboration and Knowledge Transfer}

\begin{document}

\twocolumn[
\icmltitle{Training Novices:\\The Role of Human-AI Collaboration and Knowledge Transfer}



\icmlsetsymbol{equal}{*}

\begin{icmlauthorlist}
\icmlauthor{Philipp Spitzer}{yyy}
\icmlauthor{Niklas K{\"u}hl}{yyy}
\icmlauthor{Marc Goutier}{xxx}

\end{icmlauthorlist}

\icmlaffiliation{yyy}{Karlsruhe Service Research Institute, Karlsruhe Institute of Technology, Karlsruhe, Germany}
\icmlaffiliation{xxx}{Information Systems \& E-Services, Technical University of Darmstadt, Darmstadt, Germany}

\icmlcorrespondingauthor{Philipp Spitzer}{philipp.spitzer@kit.edu}

\icmlkeywords{Machine Learning, ICML, Knowledge Retention, XAI, Human-AI Collaboration, Artificial Intelligence}

\vskip 0.3in
]



\printAffiliationsAndNotice{}  % leave blank if no need to mention equal contribution

\begin{abstract}
Across a multitude of work environments, expert knowledge is imperative for humans to conduct tasks with high performance and ensure business success. These humans possess task-specific expert knowledge (TSEK) and hence, represent subject matter experts (SMEs). However, not only demographic changes but also personnel downsizing strategies lead and will continue to lead to departures of SMEs within organizations, which constitutes the challenge of how to retain that expert knowledge and train novices to keep the competitive advantage elicited by that expert knowledge. SMEs training novices is time- and cost-intensive, which intensifies the need for alternatives. Human-AI collaboration (HAIC) poses a way out of this dilemma, facilitating alternatives to preserve expert knowledge and teach it to novices for tasks conducted by SMEs beforehand. In this workshop paper, we (1) propose a framework on how HAIC can be utilized to train novices on particular tasks, (2) illustrate the role of explicit and tacit knowledge in this training process via HAIC, and (3) outline a preliminary experiment design to assess the ability of AI systems in HAIC to act as a trainer to transfer TSEK to novices who do not possess prior TSEK.
\end{abstract}

\section{Introduction}
Decision-making and task execution in organizational activities require humans with expertise to conduct the respective tasks. Such subject matter experts (SMEs) possess task-specific expert knowledge (TSEK) and are crucial for sustaining business activities. In several domains, artificial intelligence (AI) systems are not deployed as standalone applications performing tasks or making decisions but are rather installed as agents to collaborate with SMEs. The AI system not only provides accurate predictions but can provide explanations through explainable artificial intelligence (XAI) \cite{Adadi2018}. After the AI system is trained on data provided by the SME, it accomplishes the task accordingly. This collaboration not only has the potential to improve performances in tasks where, for instance, both agents show complementarity \cite{Fuegener2021, Hemmer2022}, but also act as a trainer to teach novices new tasks. In this work, novices present humans with no prior TSEK.

Demographic changes lead to the retirement of SMEs with fewer people to replace them \cite{Engbom2017}. Additionally, downsizing strategies of organizations but also people switching jobs accelerate this hazard and jeopardize the competitive advantage of organizations since knowledge as a crucial resource of those SMEs is lost \cite{Levy2011}.

Knowledge itself can be distinguished into its explicit and tacit forms. Whereas explicit knowledge can be articulated with natural language and, for instance, stored in documentation or wikis, tacit knowledge is difficult to transfer and can only be partially expressed \cite{Nonaka1994, cavusgil2003}. This poses challenges of how to retain the knowledge of SMEs within organizations and how to transfer it to novices. While training such novices binds resources and takes time, HAIC can overcome these difficulties by utilizing the AI system as a trainer.

There is little knowledge on how AI systems in HAIC can be leveraged to train novices on a new task and therefore substitute SMEs in this training process. On top of that, this also brings up the question of whether it is possible to preserve TSEK with the aid of AI in HAIC. Thus, in our work, we examine how AI systems capture TSEK and address the following research questions:
\begin{questions}
    \item Can AI systems train novices with no prior task-specific expert knowledge on new tasks through human-AI collaboration? \label{itm:qwithlabel}
\end{questions}
To address this research question, we present a framework outlining how HAIC can be utilized to train novices and distinguish the different forms of knowledge in this process. We propose a preliminary design to examine this research question in an experiment for a future research agenda.

\begin{figure*}[ht]
\vskip 0.2in
\begin{center}
\centering{\includegraphics[width=\textwidth]{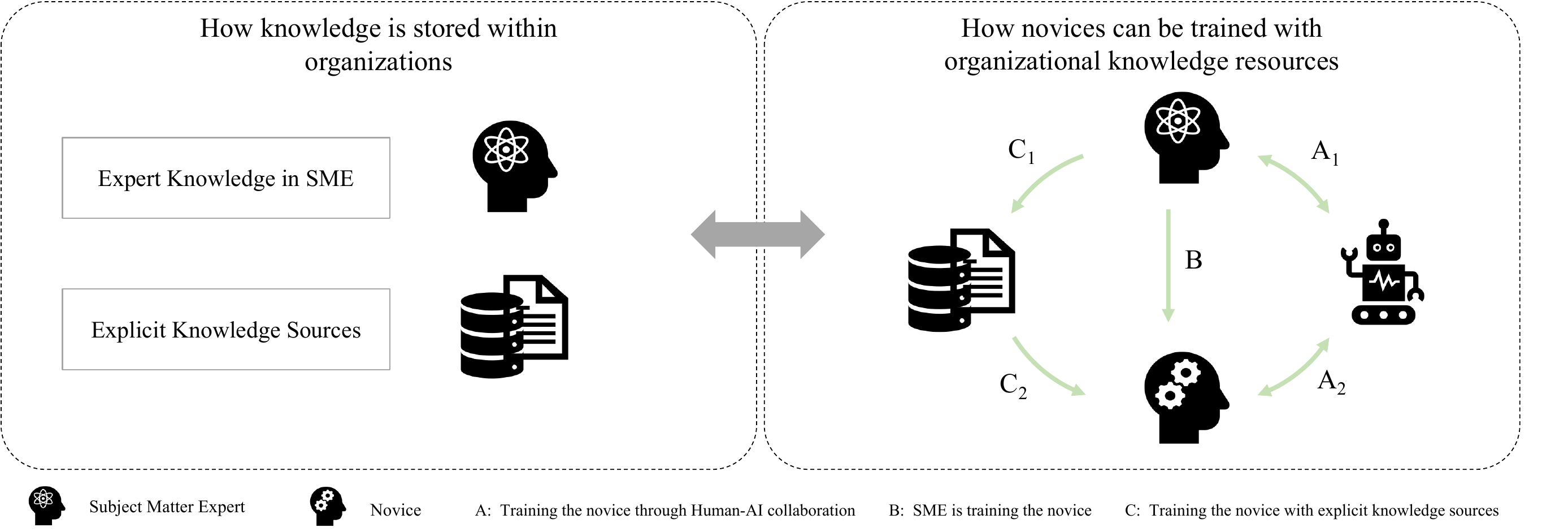}}
\caption{On the left side the conventional ways are illustrated how knowledge is stored. On the right side the different training options are displayed to teach novices new tasks within organizations. $ A_{1} $ / $ A_{2} $ illustrates the training via HAIC, $B$ via SME and $ C_{1} $ / $ C_{2} $ through explicit knowledge sources.}
\label{induction}
\end{center}
\vskip -0.2in
\end{figure*}

\section{Related Work}
\textit{Human-AI Collaboration.} Recent advances in research show promising results on the ability of AI systems to support decision-makers in their daily work to complement their tasks \cite{Bennett2013, Goldstein2018, Bullock2020, kuhl2020supporting}. Academia in the area of this socio-technical collaboration of human and AI systems focuses on XAI to enhance a human's ability to judge the predictions of these black-box systems \cite{Adadi2018, Dosilovic2018}. \citet{kuhl2020you} outline the impact of personalized explanations through XAI on task performance. Studies point out the ability to assess the AI's prediction as crucial to not forfeit overreliance and achieve high performance \cite{Bussone2015, Zhang2020}. Numerous researchers, however, refute the latter and constitute XAI can also have a negative impact on the team performance in HAIC when the AI system's prediction is inaccurate \cite{Bansal2021, Hemmer2021}. However, complementary capabilities of both agents can lead to higher overall team performance \cite{Fuegener2021, Hemmer2022}. \citet{Dellermann2019} argue that by this complementarity, through which the intelligence of both agents can be combined, AI systems and humans can learn from each other. Thereby, AI systems can teach humans new tasks since they possess unique TSEK and function as a repository of knowledge. In fact, prior work has focused on opportunities for AI systems to enhance training processes and highlights the need to incorporate AI systems as a trainer \cite{edwards2018teacher, maity2019identifying}.

\textit{Knowledge as a Crucial Resource.} The demographic transition of an aging workforce and organizations' downsizing strategies emphasize the need to retain and share expert knowledge that SMEs possess to assure ongoing business success and maintain the competitive advantage evoked through it \cite{Ambrosini2001, Levy2011, Schmitt2012, Wang2012, Burmeister2016}. Knowledge retention within organizations presents a wide research area, in which researchers focus on techniques to store and transfer the expertise of individuals. A key aspect denotes the research that investigates which knowledge forms are feasible to capture and preserve. The differentiation of knowledge into its \emph{explicit} and \emph{tacit} form is stressed by many researchers in literature and has evolved into a discourse for several decades \cite{Polanyi1962, BERRY1987, Nonaka1994}. Various researchers emphasize explicit knowledge as the form that humans can articulate with language \cite{nonaka2007knowledge, Lam2000}. \citet{Alavi2001} state that it can be captured, for instance, in a manual for a specific product. On the other hand, tacit knowledge is characterized as know-how, with which one is able to perform particular actions \cite{Ryle1945}. It represents human intuition, technical skills, or experience that cannot be expressed and only hardly transferred \cite{Nonaka1994, Lam2000, Gorman2002}. 

\citet{Hadjimichael2019} give an overview of the different positions that researchers take on the distinction between explicit and tacit knowledge. The authors argue that the interactional perspective of researchers consents to the idea that tacit knowledge can be converted into data, and used by AI systems to learn specific tasks, allowing this tacit knowledge to be transferred. Another study suggests utilizing AI systems as a medium to formalize tacit knowledge  \citep{Fenstermacher2005}. By this, TSEK is preserved within AI systems and can be transferred and used to train novices in particular tasks. Such scenarios, where not humans but AI systems can be deployed as a trainer, are noted by several researchers \cite{Dellermann2019, edwards2018teacher}.

\section{Conceptual Framework}
\subsection{HAIC for the Training of Novices}
\begin{figure*}[ht]
\vskip 0.2in
\begin{center}
\centering{\includegraphics[width=\textwidth]{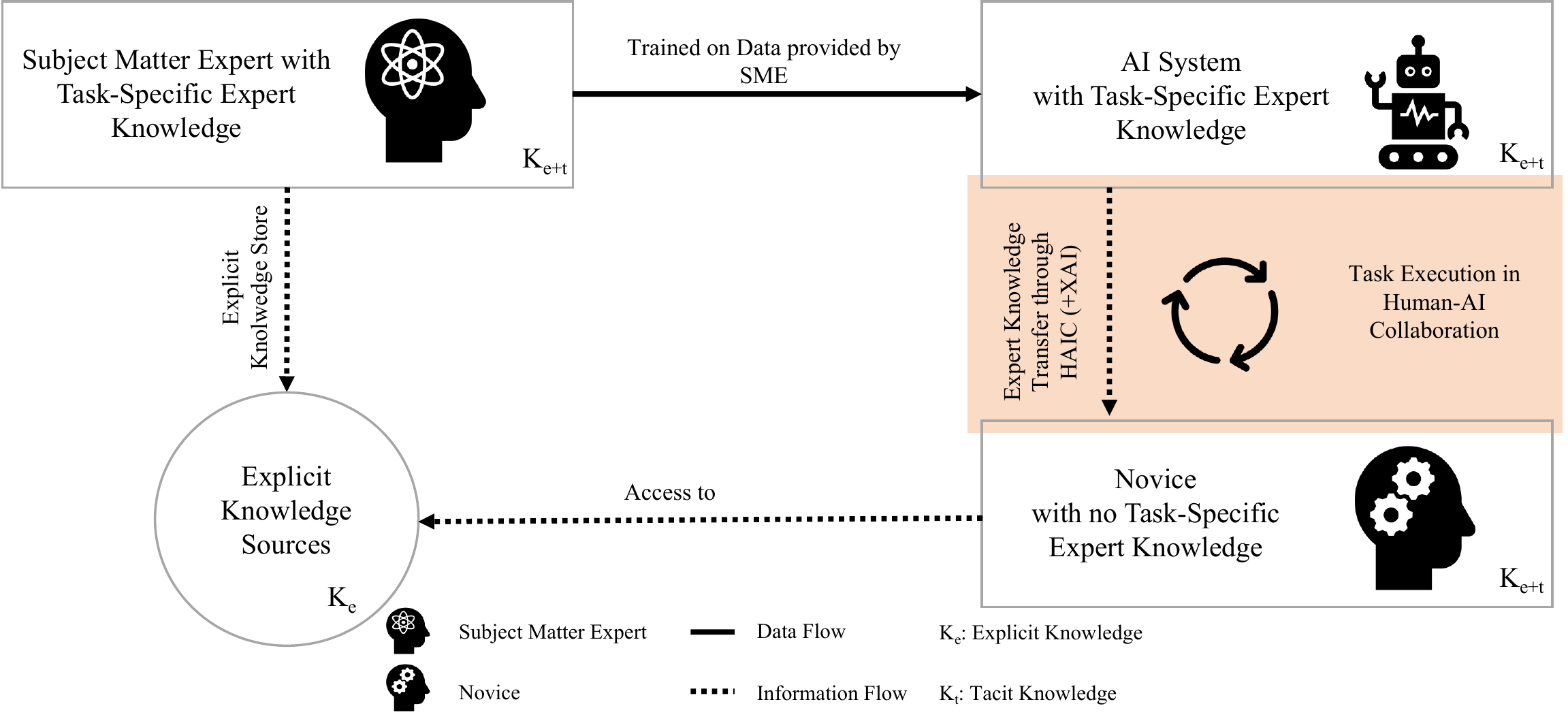}}
\caption{Conceptual Framework illustrates the training of novices through HAIC and the different forms of knowledge in this training.}
\label{knowledgeinduction}
\end{center}
\vskip -0.2in
\end{figure*}

Training of novices becomes a crucial process for organizations in the event that SMEs leave the organization. The training of novices by SMEs is costly since it takes up resources from an organizational point of view. \Cref{knowledgeinduction} illustrates how knowledge is held within organizations and how these knowledge sources are used to train novices on particular tasks.

Training novices can be accomplished in three different forms. One form represents the conventional way of training a novice through a handover phase with the SME ($B$) \citep{swanson1997training}. Here, the SME is assigned to instruct the novice and teach the execution of the task. This way, the novice can complete the task proficiently.

Another form comprises the training via explicit knowledge sources ($ C_{1} $ / $ C_{2} $). SMEs store their explicit knowledge for relevant tasks in knowledge management systems, that can be accessed by novices \citep{chau2003knowledge}. Novices can read through those explicated sources and build their TSEK based on these.

A third option depicts the training through HAIC. Here, the AI system is trained by the SME, for instance, through supervised machine learning where the SME is providing labeled data. The AI system in this collaboration has the ability to perform the particular task with sufficient performance and therefore possesses TSEK, which is illustrated as $ A_{1} $ in \Cref{knowledgeinduction}. Note that task execution can be conducted within a collaborative process where knowledge can be passed in both directions, which is why the arrow on $ A_{1} $ is pointing accordingly. Through HAIC the AI system can teach a novice how to perform this task by collaborating with the novice ($ A_{2} $). We achieve this, by the AI system not only revealing its predictions but also supplementing them with explanations of its decisions \citep{goyal2019counterfactual}. Hence, the AI system provides additional instructions and is passing on its expert knowledge through which the novice is enabled to perform the task.

\subsection{The Role of Knowledge in Training}
Organizations preserve tacit knowledge and explicit knowledge in different ways. Explicit knowledge can be articulated and, hence, stored, for instance, in documentation and wikis, which is illustrated as \textit{explicit knowledge sources} in \Cref{knowledgeinduction}. Tacit knowledge cannot be transferred in such a way \cite{Lam2000}. In this work we propose AI systems as trainers to teach novices and thereby transfer expert knowledge, specifically, explicit and tacit knowledge. Thus, with our framework, we outline how this expert knowledge transfer is conducted within HAIC to overcome the drawbacks of training a novice on a new task by an SME. We demonstrate this in a supervised machine learning setting as a subdomain of machine learning. Nevertheless, it can be applied to other domains of machine learning and AI as well.

The AI system is trained on data provided by the SME until it performs a particular task with sufficient performance and therefore possesses TSEK. It represents a knowledge store of expert knowledge for the organization. Additionally, the SME can store their explicit part of TSEK, for instance, through documentation. Whereas data is provided to the AI system for training on a particular task, the explicit knowledge store is established through the provision of information on the task by SMEs to accumulate all explicit knowledge at hand. We make use of these terms in alignment with the data, information, knowledge, and wisdom (DIKW) hierarchy of \citet{rowley2007wisdom}. Novices have access to those explicit knowledge sources but are trained by the AI system, which possesses TSEK. Across this collaboration of both agents, TSEK is transferred. Note that in this process, TSEK is not replicated but rather transferred to ensure the adequate execution of a particular task. The knowledge representation within the SME and the novice might differ, but both are enabled to execute the respective task sufficiently. Through this training, the novice becomes an SME on the particular task.

We hypothesize that training novices can be enhanced by utilizing the AI system as a trainer in HAIC. Hence, the AI system is deployed as a repository of TSEK within organizations and used to teach novices new tasks. We postulate the following hypothesis:
\begin{hyp}
Training novices with no prior knowledge of a particular task can be conducted through human-AI collaboration.
\end{hyp}
Training novices by AI systems is achieved through collaboration between both agents, in which the AI system provides its predictions along with explanations through XAI. Similar to training through SMEs, not all of the tacit knowledge may be externalized, but training with the additional information through XAI represents an alternative to transferring the TSEK. Thus, we postulate the hypothesis:
\begin{hyp}
In the training process of novices through human-AI collaboration, the AI system's prediction supplemented with explanations through XAI enhances the novice's comprehension of the task and thus, advances the training process.
\end{hyp}
In the next chapter, we describe a preliminary design for an experiment to assess and evaluate our hypotheses.

\section{Preliminary Experiment Design}
Within an experiment, we intend to test the hypotheses postulated in the previous chapter. Thus, we examine whether an AI system within HAIC is not only able to function as a repository of knowledge within organizations but also as a trainer to teach novices with no prior knowledge of a new task. Compared to other studies in this field \cite{goyal2019counterfactual, Bansal2021, schoeffer2022there}, we focus on conducting the experiment on inexperienced participants to assess the ability of AI systems within HAIC to train novices.

We test the hypotheses by using a house pricing dataset. For this regression task, other datasets are also applicable. The dataset consists of tabular data and images of the houses \cite{ted8080}. In total there are 15,474 samples. We intend to only use the tabular data for the experiment. Data is randomly sampled and split into 80\% training samples and 20\% test samples, from which 20 instances are held out to use in the experiment. A machine learning algorithm is trained on the training samples. Our experiment involves three treatments. In each treatment, participants will predict the housing prices on the 20 instances held out. After each sample they are provided the ground truth. We also measure prior knowledge and AI literacy of the participants as control variables to assure they do not possess expertise in house price prediction and to assess the comprehension of AI. 

The treatments are set up as follows:
\begin{itemize}
\item In the first treatment, participants only predict the house prices with no further assistance.
\item In the second treatment, participants will additionally get access to documentation that experts on the task created.
\item In the third treatment, participants are supported by the AI system, which reveals its prediction and gives explanations through XAI to the participants after their prediction.
\end{itemize}
Participants are presented the ground truth after each sample. We want to assess whether revealing the AI system's prediction and supplementing it with explanations through XAI leads to a TSEK transfer by providing the novice with additional information. By measuring the performance in each treatment we conceive how much TSEK the novices acquire and hence, whether training through HAIC is feasible. Through this experiment, we want to evaluate whether AI systems can be implemented as a trainer to teach novices with no prior knowledge of such a regression task and examine the impact the predictions and explanations of the AI system have on the novice's ability to learn and execute a new task. A promising field of research lies ahead.


\bibliography{paper}
\bibliographystyle{icml2022}

\end{document}